\documentclass[epj]{svjour}

\usepackage{graphics}
\usepackage{amsmath,amssymb}

\begin{document}
\title{Valence transition in the periodic Anderson model}
\author{A. H\"{u}bsch\inst{1,2} and K. W. Becker\inst{3}}
\institute{
  Department of Physics, University of California, Davis, CA 95616, USA
  \and
  Max-Planck-Institut f\"ur Physik komplexer Systeme, N\"othnitzer Stra{\ss}e
  38, 01187 Dresden, Germany
  \and
  Institut f\"{u}r Theoretische Physik, Technische Universit\"{a}t Dresden,
  D-01062 Dresden, Germany
}
\date{\today}
\abstract{
  A very rich phase diagram has recently been found in CeCu$_{2}$Si$_{2}$ from 
  high pressure experiments where, in particular, a transition between an
  intermediate valence configuration and an integral valent heavy fermion
  state has been observed. We show that such a valence transition can be
  understood in the framework of the periodic Anderson model. In particular,
  our results show a breakdown of a mixed-valence state which is accompanied
  by a drastic change in the \textit{f} occupation in agreement with
  experiment. This valence transition can possibly be interpreted as a
  collapse of the large Fermi surface of the heavy fermion state which
  incorporates not only the conduction electrons but also the localized
  \textit{f} electrons. The theoretical approach used in this paper 
  is based on the novel projector-based renormalization method (PRM). With
  respect to the periodic Anderson model, the method was before only employed
  in combination with the basic approximations of the well-known slave-boson
  mean-field theory. In this paper, the PRM treatment is performed in a more
  sophisticated manner where both mixed as well as integral valent solutions
  have been obtained. Furthermore, we argue that the presented PRM approach
  might be a promising starting point to study the competing interactions in
  CeCu$_{2}$Si$_{2}$ and related compounds. 
}
\maketitle

\section{Introduction}
\label{Intro}

Since the discovery \cite{Steglich} of a superconducting state formed by heavy
quasi-particles in CeCu$_{2}$Si$_{2}$ this and related
compounds have attracted a lot of scientific interest. Despite its long
history only recently a whole variety of new physical phases has been 
observed which was possible by the 
intriguing development of experimental techniques. 
By substituting Si by Ge in the parent
compound CeCu$_{2}$Si$_{2}$ a continuous change
from a heavy fermion (HF) superconducting phase to an antiferromagentic state
was observed \cite{Trovarelli}. An even more complex phase 
diagram has been found in pure 
CeCu$_{2}$Si$_{2}$ by applying high pressure
\cite{Yuan,Holmes}: There two superconducting phases with different pairing
mechanisms have been found besides an antiferromagnetic and a HF
phase. Furthermore, a transition between intermediate and integral valence
states has been observed. (For a recent review on superconductivity in Ce
based HF materials see Ref.~\cite{Thalmeier}.) 

\bigskip
From the theoretical point of view the periodic Anderson model (PAM) is
considered to be the basic microscopic model
for the investigation of HF systems \cite{Lee}. The PAM 
describes the interaction between localized, 
strongly correlated \textit{f} and itinerant conduction
electrons. In the limit of infinitely large Coulomb repulsion on \textit{f}
sites the PAM can be written as 
\begin{eqnarray}
  \label{G1}
  \mathcal{H} &=& \mathcal{H}_{0} + \mathcal{H}_{1} ,
\end{eqnarray}
\begin{eqnarray}
  \mathcal{H}_0 & = & \varepsilon_{f} \sum _{i,m} \hat{f}^{\dagger} _{im}
  \hat{f} _{im}
  + \sum _{{\bf k},m} \varepsilon_{{\bf k}} \ c^{\dagger}_{{\bf k}m}
  c_{{\bf k}m} , \nonumber\\
  \mathcal{H}_1 & = & \frac{1}{\sqrt{N}} \sum _{{\bf k},i,m} V_{{\bf k}}
  \left(
    \hat{f}^{\dagger} _{im} c_{{\bf k}m} \,
    e^{{\rm i}{\bf k}{\bf R}_i} + {\rm h.c.}
  \right ) .
  \nonumber
\end{eqnarray}
Here, $\varepsilon_{f}$ and $\varepsilon_{\mathbf{k}}$, both measured from the
chemical potential, are the excitation energies of localized \textit{f} and
itinerant conduction electrons. As a simplification, often 
both types of electrons are assumed to have the same angular 
momentum index $m$ with
$\nu_{f}$ values, $m=1 ... \nu_{f}$. The infinitely large local Coulomb
repulsion is taken into account by Hubbard operators 
\begin{eqnarray*}
  \hat{f}^{\dagger} _{im} &=& f^{\dagger} _{im} \prod_{\tilde m (\ne m)}
  (1- f_{i\tilde{m}}^\dagger f_{i\tilde{m}})
\end{eqnarray*}
which enable either empty or singly occupied \textit{f} sites.

Due to the complexity of the PAM, most theoretical studies only focus on
certain aspects of the rich phase diagrams of rare earth materials. Slave-boson
mean-field (SB) methods, large-N expansions, and the dynamical mean-field
theory \cite{RKKY} have been applied to discuss the interplay between 
RKKY and Kondo interactions. Thereby, a transition
between an antiferromagnetic phase and a paramagnetic state 
was discussed. On the other hand, to describe the valence 
transition and HF superconductivity in CeCu$_{2}$Si$_{2}$ 
an extended PAM was studied. This model includes an additional Cou\-lomb 
interaction between \textit{f} and
conduction electrons and was discussed within a slave-boson fluctuation
approximation \cite{Holmes,Onishi}. 

\bigskip
In this paper we apply a novel projector-based renormalization method
(PRM) \cite{Becker} to the PAM with the aim  to address the question whether a
valence transition, as experimentally 
observed in CeCu$_{2}$Si$_{2}$ \cite{Yuan,Holmes}, can occur in the plain 
model. For that purpose we extend in this paper our previous work on the PAM
\cite{Hubsch}, which was restricted to the HF phase. The PRM provides a
natural way to discuss the interplay of competing interactions which naturally
emerge from the renormalization treatment of the PAM. Therefore, we believe
that the PRM represents a suited approach for a deeper understanding of the
rich phase diagram of CeCu$_{2}$Si$_{2}$ or of related compounds. However, in
this paper we concentrate on the valence transition, nevertheless, we are able
to sketch how superconducting phases and RKKY interactions could also be
included in our approach.

\bigskip
This paper is organized as follows. In the next section we briefly describe
the novel PRM approach \cite{Becker} that is applied to the PAM in
Sec.~\ref{RenPAM}. Here, the Hubbard operators, introduced to take into
account the infinitely large Coulomb repulsion on \textit{f} sites, cause the
main problems of any theoretical approach. It will turn out that the
well-known SB theory \cite{Coleman,Fulde} as well as
our recent analytical approach based on the PRM \cite{Hubsch} do not
sufficiently prevent from unphysical states with doubly occupied \textit{f}
sites. In contrast, the modified PRM treatment of Sec.~\ref{RenPAM} strictly
suppresses doubly occupied \textit{f} sites by taking into account electronic
correlations by means of the Hubbard operators. Results are presented in
Sec.~\ref{Results} where mixed valent as well as integral valent states are
found, and a valence transition is observed. Furthermore, we compare our
results with the solutions of the SB theory and our PRM approach of
Ref.~\cite{Hubsch}. Finally, we summarize in Sec.~\ref{Summ}.

\section{Methodology}
\label{Methode}

The PRM approach \cite{Becker} starts from a decomposition of a given
many-particle Hamiltonian,
$
  \mathcal{H} = \mathcal{H}_{0} + \mathcal{H}_{1}
$,
where the perturbation $\mathcal{H}_{1}$ should not contain any terms that
commute with the unperturbed part $\mathcal{H}_{0}$. Thus, $\mathcal{H}_{1}$
represents transitions between eigenstates of $\mathcal{H}_{0}$ with
\textit{different} eigenenergies. In the following, we assume that the
eigenvalue problem of $\mathcal{H}_{0}$ is solved, 
\begin{eqnarray}
  \mathcal{H}_{0} | n^{(0)} \rangle &=& E_{n}^{(0)} | n^{(0)} \rangle.
  \nonumber
\end{eqnarray}
A crucial idea of the PRM is the definition of projection operators by
\begin{eqnarray}
  \label{G2}
  \mathbf{P}_{\lambda} {\mathcal{A}} &=&
  \sum_{m,n} | n^{(0)} \rangle \langle m^{(0)} |
  \langle n^{(0)} | \mathcal{A} | m^{(0)}\rangle \\[-1ex]
  &&\quad\times\Theta(\lambda -|E_n^{(0)}-E_m^{(0)}|). \nonumber
\end{eqnarray}
Note that ${\bf P}_{\lambda}$ and 
${\bf Q}_{\lambda} = {\bf 1} - {\bf P}_\lambda$ 
are super-operators acting on ordinary operators ${\cal A}$ of the unitary
space. $\mathbf{P}_{\lambda}$ projects on those parts of
${\cal A}$ which are formed by transition operators
$| n^{(0)}\rangle \langle m^{(0)} |$ with energy
differences $| E_n^{(0)} - E_m^{(0)} |$ less than a given
cutoff $\lambda$. ($\lambda$ is smaller than the cutoff $\Lambda$ of the
original model.) On the other hand, ${\bf Q}_{\lambda}$ projects on the
high-energy transitions of an operator. Note, in particular, that in
Eq.~\eqref{G2} neither $| n^{(0)}\rangle$ nor $| m^{(0)} \rangle$
have to be low-energy eigenstates of ${\cal H}_0$.

Next, an effective Hamiltonian $\mathcal{H}_{\lambda}$ is derived from the
original Hamiltonian $\mathcal{H}$ by an unitary transformation
\begin{eqnarray}
  \label{G3}
  {\cal H}_\lambda &=& e^{X_\lambda}\; {\cal H}\;  e^{-X_\lambda} \; ,
\end{eqnarray}
where the anti-Hermitian generator of the transformation, 
$X_{\lambda} = -X^{\dagger}_{\lambda}$, shall be chosen in such a way that 
only transition operators (between eigenstates of $\mathcal{H}_{0}$) with
transition energies less than the given cutoff $\lambda$ contribute to
$\mathcal{H}_{\lambda}$. Thus, the condition
\begin{eqnarray}
  \label{G4}
  \mathbf{Q}_{\lambda} \mathcal{H}_{\lambda} &=& 0
\end{eqnarray}
must be fulfilled and will be used below to determine $X_{\lambda}$. Note that
it is straightforward to evaluate Eqs.~\eqref{G3} and \eqref{G4} in
perturbation theory \cite{Becker}. 
However, using an appropriate ansatz for the generator
$X_{\lambda}$, the effective Hamiltonian ${\cal H}_\lambda$ can also be
calculated in non-perturbative manner. 

A renormalization scheme can be derived if the elimination procedure for the
interaction $\mathcal{H}_{1}$ is not performed in one step but rather a
sequence of unitary transformations of the form 
\begin{eqnarray}
  \label{G5}
  \mathcal{H}_{(\lambda-\Delta\lambda)} =
  e^{X_{\lambda,\Delta\lambda}}
  \, \mathcal{H}_{\lambda} \,
  e^{-X_{\lambda,\Delta\lambda}}
\end{eqnarray}
is applied to the original Hamiltonian $\mathcal{H}$. Thus, transitions
between eigenstates of $\mathcal{H}_0$ caused by the interaction
$\mathcal{H}_{1}$ are eliminated in steps where the respective
transition  energies are used as renormalization parameter
$\lambda$. Furthermore, 
\begin{eqnarray}
  \label{G6}
  \mathbf{Q}_{(\lambda-\Delta\lambda)} 
  \mathcal{H}_{(\lambda-\Delta\lambda)} &=& 0
\end{eqnarray}
is used to specify the generator $X_{\lambda,\Delta\lambda}$ of the unitary
transformation. Note that Eqs.~\eqref{G5} and \eqref{G6} describe a
renormalization step that decreases the cutoff of the Hamiltonian from
$\lambda$ to $(\lambda - \Delta\lambda)$, as one can see from a comparison
with Eqs.~\eqref{G3} and \eqref{G4}. Therefore, difference equations for the
$\lambda$ dependence of the Hamiltonian can be derived from \eqref{G5} and
\eqref{G6}, and we call the resulting equations for the parameters of the
Hamiltonian renormalization equations. Note, that the solutions of these
renormalization equations strongly depend on the parameters of the original
Hamiltonian $\mathcal{H}$, and that the limit $\lambda\rightarrow 0$ provides
the desired effective Hamiltonian without any interactions.

\section{Renormalization of the PAM}
\label{RenPAM}

In the following, we want to apply the framework of the PRM as discussed above
to the PAM. It is well known that much of the physics of the PAM  
\eqref{G1} can be understood in terms of an uncorrelated model, that is 
for vanishing  Coulomb repulsion on \textit{f} sites where the Hubbard
operators $\hat{f}^{\dagger}_{im}$ are replaced by usual fermionic operators 
$f^{\dagger}_{im}$. This model can be solved exactly. However, the parameters
have to be  renormalized appropriately. Various theoretical methods have been
developed to generate renormalized Hamiltonians. Most popular is the one
derived from slave-boson mean-field (SB) theory \cite{Coleman,Fulde}. 
Note however that only HF type solutions can be  obtained in this way. 
In particular, the SB solution breaks down if the original 
\textit{f} level is located too far below the Fermi level \cite{Franco}  
or, equivalently, if the hybridization strength between 
\textit{f} and conduction electrons becomes too weak. 

In the Hamiltonian \eqref{G1} the Hubbard operators $\hat{f}^{\dagger}_{im}$
take care of the infinitely large local Coulomb repulsion on \textit{f}
sites so that multiple occupied \textit{f} sites are strictly forbidden. 
Consequently, any effective model has to satisfy this requirement as
well. However, SB mean-field theory as well as our recent PRM treatment of the
PAM \cite{Hubsch} map the Hamiltonian of Eq.~\eqref{G1} onto an effectively
free system consisting of two non-interacting fermionic quasi-particles, 
\begin{eqnarray}
  \label{G7} 
  \mathcal{H}_{\mathrm{eff}} & = & 
  \sum _{{\bf k},m} 
  \omega_{\mathbf{k}}^{c} \ c^{\dagger}_{{\bf k}m} c_{{\bf k}m}  + 
  \sum _{\mathbf{k},m} \omega_{\mathbf{k}}^{f} \
  f^{\dagger} _{\mathbf{k}m} f_{\mathbf{k}m} + E_\mathrm{shift},
\end{eqnarray}
with renormalized parameters $\omega_{\mathbf{k}}^{c}$,
$\omega_{\mathbf{k}}^{f}$, $E_\mathrm{shift}$. It is important to notice that
due to construction the effective Hamiltonian 
$\mathcal{H}_{\mathrm{eff}}$ of Eq.\eqref{G7} \textit{does not prevent} 
from multiple occupation of \textit{f} sites. This follows from the occurrence
of the fermionic operators $f_{{\bf k}m}$ and $f_{{\bf k}m}^\dagger$ in
Eq.\eqref{G7} instead of the Hubbard operators $\hat{f}_{{\bf k}m}$ and 
$\hat{f}_{{\bf k}m}^\dagger$. However, an approximation that involves a
replacement of the Hubbard operators $\hat{f}_{{\bf k}m}$ and 
$\hat{f}_{{\bf k}m}^\dagger$ by usual fermionic operators $f_{{\bf k}m}$ and
$f_{{\bf k}m}^\dagger$ might lead to useful results as long as only very few
$f$ type states are below the Fermi level. Thus, only HF-like solutions
with a renormalized $f$ level above the Fermi level can be obtained based on
effective Hamiltonians of type \eqref{G7}, and SB mean-field theory as well as
our recent PRM treatment of the PAM \cite{Hubsch} can not describe integral
valent states.

\subsection{Renormalization ansatz} 

In this paper we want to describe the transition of the PAM between mixed
valent and integral valent states. Thus, a theoretical treatment is needed that
reliably prevents from unphysical multiple occupation of $f$ sites. For that
purpose, we again apply the framework of the PRM to the PAM, but, in contrast
to our recent work \cite{Hubsch}, we now keep the Hubbard operators during the
whole renormalization procedure. Thus, the renormalization ansatz reads
\begin{eqnarray}
  \label{G8}
  \mathcal{H}_{\lambda} & = & 
  \mathcal{H}_{0,\lambda} + \mathcal{H}_{1,\lambda} \\[1ex]
  \mathcal{H}_{0,\lambda} & = &
  \mu_{f,\lambda} \sum_{\mathbf{k},m} 
  \hat{f}^{\dagger}_{\mathbf{k}m} \hat{f}_{\mathbf{k}m} +  
  \sum_{\mathbf{k},m} \Delta_{\mathbf{k},\lambda} 
  \left(
    \hat{f}^{\dagger} _{\mathbf{k}m} \hat{f}_{\mathbf{k}m} 
  \right)_{\mathrm{NL}} 
  \nonumber \\
  &&
  + \sum _{{\bf k},m} 
  \varepsilon_{{\bf k},\lambda} \ c^{\dagger}_{{\bf k}m} c_{{\bf k}m}  + 
  E_{\lambda} \nonumber \\[1ex]
  \mathcal{H}_{1,\lambda} & = &
  \mathbf{P}_{\lambda} \mathcal{H}_{1} \,=\,
  \sum_{\mathbf{k},m} V_{\mathbf{k}} \ 
  \mathbf{P}_{\lambda}\left(
    \hat{f}^{\dagger}_{\mathbf{k}m} c_{{\bf k}m} + \mathrm{h.c.} 
  \right)
  \nonumber
\end{eqnarray}
after all excitations with energies larger than the cut-off $\lambda$ have
been eliminated. Due to the renormalization process all parameters depend on
$\lambda$, and an additional energy shift $E_{\lambda}$ and a hopping between
different $f$ sites,
\begin{eqnarray*}
  \left(
    \hat{f}^{\dagger} _{\mathbf{k}m} \hat{f}_{\mathbf{k}m} 
  \right)_{\mathrm{NL}} 
  & = &
  \frac{1}{N}\sum_{i, j(\not{=}i)}
  \hat{f}_{im}^{\dagger} \hat{f}_{jm} 
  e^{i\mathbf{k}(\mathbf{R}_{i}-\mathbf{R}_{j})}, 
\end{eqnarray*}
have been generated. Here, we have introduced Fourier transformed $f$
operators,
\begin{eqnarray*}
  \hat{f}^{\dagger}_{\mathbf{k}m} &=&
  \frac{1}{\sqrt{N}} 
  \sum_{i} \hat{f}^{\dagger}_{im} e^{i\mathbf{k}\cdot\mathbf{R}_{i}}.
\end{eqnarray*}
The initial parameter values of the original model (at cut-off 
$\lambda = \Lambda$) are
\begin{eqnarray}
  \label{G9}
  \mu_{f,\Lambda} = \varepsilon_{f}, \;
  \Delta_{\mathbf{k},\Lambda} = 0, \;
  \varepsilon_{{\bf k},\Lambda} = \varepsilon_{{\bf k}}, \;
  E_{\Lambda} = 0.
\end{eqnarray}

To perform the PRM scheme we also need the commutator of the unperturbed
Hamiltonian with the hybridization. For convenience, we introduce the
unperturbed Liouville operator $\mathbf{L}_{0,\lambda}$ which is defined by 
$\mathbf{L}_{0,\lambda}\mathcal{A} =[\mathcal{H}_{0,\lambda}, \mathcal{A}]$
for any operator variable $\mathcal{A}$, and to simplify the calculations, the
one-particle operators $\hat{f}^{\dagger}_{\mathbf{k}m}$ and
$c^{\dagger}_{\mathbf{k}m}$ are considered as approximate eigenoperators of
$\mathbf{L}_{0,\lambda}$, 
\begin{eqnarray}
  \label{G10}
  \mathbf{L}_{0,\lambda} \ \hat{f}^{\dagger} _{\mathbf{k}m} c_{\mathbf{k}m}
  &\approx&
  \left( 
    \varepsilon_{f,\lambda} + D\Delta_{\mathbf{k},\lambda} - 
    \varepsilon_{\mathbf{k},\lambda} 
  \right)
  \hat{f}^{\dagger}_{\mathbf{k}m} c_{\mathbf{k}m}.
\end{eqnarray}
Here, we introduced the local $f$ energy,
\begin{eqnarray}
  \label{G11}
  \varepsilon_{f,\lambda}  & = &
  \mu_{f,\lambda} - D \bar{\Delta}_{\lambda} ,
\end{eqnarray}
and defined 
$
  D = 1 - \langle \hat{n}_{i}^{f} \rangle + 
  \langle \hat{n}_{i}^{f} \rangle / \nu_{f}
$
and 
$
  \bar{\Delta}_{\lambda} = 
  \frac{1}{N}\sum_{\mathbf{k}}\Delta_{\mathbf{k},\lambda}
$.
The factors $D$ in Eqs.~\eqref{G9} and \eqref{G10} are caused by the Hubbard
operators in the renormalization ansatz \eqref{G8}. Similar expressions
without factors $D$ have also been found in Ref.~\cite{Hubsch} where a
renormalization ansatz consisting of fermionic quasi-particles has been
used.

As one can see from Eq.~\eqref{G10}, the operator product 
$\hat{f}^{\dagger}_{\mathbf{k}m} c_{\mathbf{k}m}$ can also be interpreted as an
approximate eigenoperator of the Liouville operator
$\mathbf{L}_{0,\lambda}$. The corresponding eigenvalues are excitation
energies and can be used to rewrite $\mathcal{H}_{1,\lambda}$,
\begin{eqnarray*}
  \mathcal{H}_{1,\lambda} &=& 
  \sum_{\mathbf{k},m} 
  \Theta_{\mathbf{k},\lambda} \ V_{\mathbf{k}} 
  \left(
    \hat{f}^{\dagger}_{\mathbf{k}m} c_{{\bf k}m} + \mathrm{h.c.} 
  \right),
\end{eqnarray*}
where the $\Theta$ functions 
\begin{eqnarray*}
  \Theta_{\mathbf{k},\lambda} &=&
  \Theta
  \left(
    \lambda - 
    \left|  
      \varepsilon_{f,\lambda} + D\Delta_{\mathbf{k},\lambda} - 
      \varepsilon_{\mathbf{k},\lambda}
    \right|
  \right)
\end{eqnarray*}
restrict the particle-hole excitations to transition energies smaller than
$\lambda$.

\subsection{Renormalization equations} 

Next we want to follow the discussion of Ref.~\cite{Hubsch} to derive
renormalization equations for the parameters of the renormalized Hamiltonian 
$\mathcal{H}_{\lambda}$. It turns out that the actual calculations are
only slightly modified by the new renormalization ansatz \eqref{G8} which
now includes correlated Hubbard operators.

To evaluate the new Hamiltonian $\mathcal{H}_{(\lambda-\Delta\lambda)}$
according to Eq.~\eqref{G5}, an unitary transformation has to be performed to
eliminate excitations within the energy shell between
$(\lambda-\Delta\lambda)$ and $\lambda$. As in Ref.~\cite{Hubsch}, we use the
following operator ansatz for the generator 
$X_{\lambda,\Delta\lambda}$ of the unitary transformation, 
\begin{eqnarray*}
  X_{\lambda, \Delta \lambda} =
  \sum_{\mathbf{k},m}
  \Theta_{\mathbf{k}}(\lambda, \Delta\lambda) \,
  A_{\mathbf{k}}(\lambda, \Delta\lambda) \,
  ( \hat{f}_{\mathbf{k}m}^{\dagger} c_{\mathbf{k}m} -
  c_{\mathbf{k}m}^{\dagger} \hat{f}_{\mathbf{k}m})
\end{eqnarray*}
where the $\Theta_{\mathbf{k}}(\lambda, \Delta\lambda)$ are products of two 
$\Theta$ functions,
\begin{eqnarray*}
  \Theta_{\mathbf{k}}(\lambda, \Delta\lambda) &=&
  \Theta_{\mathbf{k},\lambda} 
  \left[ 1 - \Theta_{\mathbf{k},(\lambda - \Delta\lambda)} \right].
\end{eqnarray*}
Note that the $\Theta_{\mathbf{k}}(\lambda, \Delta\lambda)$ confine the
excitations which have to be eliminated by the renormalization step from
$\lambda$ to $(\lambda - \Delta\lambda)$. The unknown parameters
$A_{\mathbf{k}}(\lambda, \Delta\lambda)$ have to be fixed in such a way so
that only transition with energies smaller than the new 
cut-off $(\lambda-\Delta\lambda)$ contribute to
$\mathcal{H}_{(\lambda-\Delta\lambda)}$. 

\bigskip
As described in Ref.~\cite{Hubsch}, equations for the parameters
$A_{\mathbf{k}}(\lambda, \Delta\lambda)$ of the generator of the unitary
transformation as well as for the parameters of the
renormalized Hamiltonian $\mathcal{H}_{\lambda}$ can be found by comparing the
coefficients of the operators in the renormalization ansatz \eqref{G8} at
cutoff $(\lambda - \Delta\lambda)$ and in the explicitly evaluated
unitary transformation \eqref{G5}. 

Thus, we obtain the following equations:
\begin{eqnarray}
  \label{G12}
  \lefteqn{A_{\mathbf{k}}(\lambda, \Delta\lambda) \, = \,}
  && \\[1ex]
  &=&
  \frac{\Theta_{\mathbf{k}}(\lambda, \Delta\lambda)}{2\sqrt{D}}
  \, \arctan \left[
    \frac{2 \sqrt{D} V_{\mathbf{k}}
    }{
      \mu_{f,\lambda} +
      D \left( \Delta_{\mathbf{k},\lambda} - \bar{\Delta}_{\lambda} \right) -
      \varepsilon_{\mathbf{k},\lambda}
    }
  \right]\nonumber
\end{eqnarray}
\begin{eqnarray}
  \label{G13}
  \lefteqn{
    \varepsilon_{\mathbf{k},(\lambda-\Delta\lambda)} -
    \varepsilon_{\mathbf{k},\lambda}
    \, = \,
  } && \\[1ex]
  &=&
  - \, \frac{1}{2}
  \left[
    \mu_{f,\lambda} +
    D \left( \Delta_{\mathbf{k},\lambda} - \bar{\Delta}_{\lambda} \right) -
    \varepsilon_{\mathbf{k},\lambda}
  \right] \nonumber \\
  && \qquad \times
  \left\{
    \cos \left[ 2 \sqrt{D} A_{\mathbf{k}}(\lambda, \Delta\lambda) \right] - 1
  \right\} \nonumber \\
  &&
  - \,
  \sqrt{D} V_{\mathbf{k}}
  \sin \left[ 2 \sqrt{D} A_{\mathbf{k}}(\lambda, \Delta\lambda) \right] ,
  \nonumber
\end{eqnarray}
\begin{eqnarray}
  \label{G14}
  \Delta_{\mathbf{k},(\lambda-\Delta\lambda)} - \Delta_{\mathbf{k},\lambda}
  &=&
  - \frac{1}{D} \,
  \left[
    \varepsilon_{\mathbf{k},(\lambda-\Delta\lambda)} -
    \varepsilon_{\mathbf{k},\lambda}
  \right]
\end{eqnarray}
\begin{eqnarray}
  \label{G15}
  \lefteqn{
    \mu_{f,(\lambda-\Delta\lambda)} - \mu_{f,\lambda}
    \, = \,
  } && \\[1ex]
  &=&
  - \, \frac{1}{D} \frac{1}{N} \sum_{\mathbf{k}}
  \left[
    \varepsilon_{\mathbf{k},(\lambda-\Delta\lambda)} -
    \varepsilon_{\mathbf{k},\lambda}
  \right] \nonumber \\
  &&
  \qquad \qquad \times
  \left[
    1 + (\nu_f -1)
    \left\langle c^{\dagger}_{{\bf k}m} c_{{\bf k}m} \right\rangle
  \right] \nonumber \\
  &&
  +\, \frac{\nu_{f} - 1}{4 D^{3/2}} \, \frac{1}{N}
  \sum_{\mathbf{k}}
  \left\{
    \left[
      \mu_{f,\lambda} +
      D \left( \Delta_{\mathbf{k},\lambda} - \bar{\Delta}_{\lambda} \right) -
      \varepsilon_{\mathbf{k},\lambda}
    \right]
  \right. \nonumber \\
  &&
  \qquad\qquad \times
  \sin\left[2 \sqrt{D} A_{\mathbf{k}}(\lambda, \Delta\lambda) \right]
  \nonumber\\
  && 
  \qquad
  -\,
  \left.
    2 \sqrt{D} V_{\mathbf{k}}
    \left\{
      \cos \left[2 \sqrt{D} A_{\mathbf{k}}(\lambda, \Delta\lambda) \right] - 1
    \right\}
  \right\}
  \nonumber \\
  &&
  \qquad\qquad
  \times\left\langle
    \hat{f}_{\mathbf{k}m}^{\dagger}c_{\mathbf{k}m}+\mathrm{h.c.}
  \right\rangle \nonumber\\
  &&
  - \,
  \frac{\nu_{f} - 1}{2D} \, \frac{1}{N}
  \sum_{\mathbf{k}}
  \left[
    \mu_{f,\lambda} -
    D \left( \Delta_{\mathbf{k},\lambda} - \bar{\Delta}_{\lambda} \right) -
    \varepsilon_{\mathbf{k},\lambda}
  \right]
  \nonumber \\
  && \qquad \qquad
  \times \, A_{\mathbf{k}}(\lambda,\Delta\lambda)
  \left\langle
    \hat{f}_{\mathbf{k}m}^{\dagger}c_{\mathbf{k}m}+\mathrm{h.c.}
  \right\rangle \nonumber
\end{eqnarray}
\begin{eqnarray}
  \label{G16}
  \lefteqn{E_{(\lambda-\Delta\lambda)} - E_{\lambda} \, = \,} && \\[1ex]
  &=&
  -\, N \langle \hat{n}_{i}^{f} \rangle
  \left[
    \mu_{f,(\lambda-\Delta\lambda)} - \mu_{f,\lambda}
  \right] 
  \nonumber \\
  &&
  \qquad
  -\,
  \frac{\langle \hat{n}_{i}^{f} \rangle}{D} 
  \sum_{\mathbf{k}}
  \left[
    \varepsilon_{\mathbf{k},(\lambda-\Delta\lambda)} -
    \varepsilon_{\mathbf{k},\lambda}
  \right] \nonumber
\end{eqnarray}
Note that besides the factor $1/D$ in Eq.~\eqref{G14} these renormalization
equations exactly agree with those derived in Ref.~\cite{Hubsch}. However, the
underlying Hamiltonians differ significantly because now the renormalization
ansatz \eqref{G8} contains correlation effects by means of the Hubbard
operators. It will turn out that the Hubbard operators not only complicate the
further evaluation of the renormalization equations but also successfully
prevent the system from unphysical multiple occupation of the $f$ sites. 

In deriving the renormalization equations \eqref{G12}-\eqref{G16} a
factorization approximation has been employed so that the obtained equations
still depend on expectation values which have to be determined simultaneously
(see Ref.~\cite{Hubsch} for details). Furthermore, an expansion in $1/\nu_{f}$
has been avoided (and spin fluctuations have been neglected) so that the
derived renormalization equations are valid for large as well as small
degeneracies $\nu_{f}$. The limit $\lambda \rightarrow 0$ provides the
parameters $\tilde{\varepsilon}_{{\bf k}}$, $\tilde{\mu}_{f}$,
$\tilde{\Delta}_{\mathbf{k}}$, and $\tilde{E}$ of the effective Hamiltonian 
$
  \tilde{\mathcal{H}} = \mathcal{H}_{\lambda\rightarrow 0}
  = \mathcal{H}_{0, \lambda\rightarrow 0}
$,
\begin{eqnarray}
  \label{G17}
  \tilde{\mathcal{H}} & = & 
  \sum _{{\bf k},m} 
  \tilde{\varepsilon}_{{\bf k}} \ c^{\dagger}_{{\bf k}m} c_{{\bf k}m}  + 
  \tilde{\mu}_{f} \sum _{\mathbf{k},m} 
  \hat{f}^{\dagger}_{\mathbf{k}m} \hat{f}_{\mathbf{k}m} \\
  &&
  + \sum _{{\bf k},m} \tilde{\Delta}_{\mathbf{k}} 
  \left(
    \hat{f}^{\dagger} _{\mathbf{k}m} \hat{f}_{\mathbf{k}m} 
  \right)_{\mathrm{NL}}
  + \tilde{E}\, , \nonumber
\end{eqnarray}
we are interested in. Here, it is important to notice that the renormalized
Hamiltonian $\tilde{\mathcal{H}}$ no longer contains the hybridization between
conduction and localized electrons. However, $\tilde{\mathcal{H}}$ is
\textit{not} a non-interacting fermionic system because $\tilde{\mathcal{H}}$
still takes into account electronic correlations by means of the Hubbard
operators $\hat{f}^{\dagger}_{\mathbf{k}m}$. Note that these correlations turn
out to be crucial for a description of integral valent states. On the other
hand, the Hubbard operators $\hat{f}^{\dagger}_{\mathbf{k}m}$ also cause
challenging difficulties in the further theoretical treatment because they do
\textit{not} obey the usual fermionic anticommutator relations.

\subsection{Approximate solutions} 
\label{Approx_Sol}

In the following we want to develop a strategy to solve the renormalization
equations \eqref{G12}-\eqref{G16} approximately. Here, similar approximations
as in Ref.~\cite{Hubsch} shall be used to decouple the renormalization of the
different $\mathbf{k}$ values. In this way, all relevant quantities can be
expressed as functions of a renormalized $f$ energy $\tilde{\varepsilon}_{f}$
which is determined by numerical minimization of the free energy.  

\bigskip
As in Ref.~\cite{Hubsch}, we use the following approximations for further
evaluation of the renormalization equations \eqref{G12}-\eqref{G16}:
\begin{description}
  \item[(i)]
  All expectation values (which occur due to the exploited factorization
  approximation) are assumed to be independent from the renormalization 
  parameter $\lambda$ and are calculated using the full Hamiltonian
  $\mathcal{H}$. 
  \item[(ii)]
  To decouple the renormalization of the different $\mathbf{k}$ values, the 
  $\lambda$ dependence of the renormalized \textit{f} level is neglected, 
  $\mu_{f,\lambda} - D\bar{\Delta}_{\lambda} \approx \tilde{\varepsilon}_{f}$.
  The spirit of this approximation is similar to that assumed in the
  SB theory where a renormalized \textit{f} energy is also used from the very
  beginning. Note that $\tilde{\varepsilon}_{f}$ has to be interpreted as
  local $f$ energy of the renormalized model \eqref{G17}.
\end{description}

\bigskip
At this point it is important to notice, that our old analytical solution of
Ref.~\cite{Hubsch} can be easily obtained if the Hubbard operators in the
final Hamiltonian \eqref{G17} are replaced by usual fermionic
operators. Formally, one employs
\begin{description}
  \item[(iii)]
  $
    \sum _{\mathbf{k},m} \hat{f}^{\dagger}_{\mathbf{k}m} \hat{f}_{\mathbf{k}m}
    \approx
    \sum _{\mathbf{k},m} f^{\dagger}_{\mathbf{k}m} f_{\mathbf{k}m}
  $
  and \\[1ex]
  $
    ( \hat{f}^{\dagger} _{\mathbf{k}m} \hat{f}_{\mathbf{k}m} )_{\mathrm{NL}}
    \approx
    D \ ( f^{\dagger}_{\mathbf{k}m} f_{\mathbf{k}m} )_{\mathrm{NL}}
  $.
\end{description}
to ensure that, on a mean-field level, the renormalized Hamiltonian does not
generate unphysical states. However, as already discussed above, the
obtained effective model does not prevent anymore from multiple occupation of
$f$ sites if (iii) has been employed. We have already argued that such an
approximation can only lead to useful results as long as only very view $f$
type states below the Fermi level are occupied. Thus, only HF-like solutions
can be observed in this way. To obtain the analytical solution of
Ref.~\cite{Hubsch}, one also has to employ
\begin{description}
  \item[(iv)]
  $
    \frac{1}{N}\sum_{\mathbf{k}} \Delta_{\mathbf{k},\lambda} \approx 
    \tilde{\Delta} \approx 0
  $ 
\end{description}
for further simplification.

\bigskip
In the following we only want to employ approximations (i) and (ii). In
particular, we keep the Hubbard operators in the final Hamiltonian \eqref{G17}
so that both mixed valent and integral valent states can be described.

Eqs.~\eqref{G14} and \eqref{G16} can be easily integrated between the lower
cutoff $\lambda\rightarrow 0$ and the cutoff of the original model $\Lambda$, 
\begin{eqnarray}
  \label{G18}
  \tilde{\Delta}_{\mathbf{k}} &=&
  - \, \frac{1}{D} 
  \left[ \tilde{\varepsilon}_{\mathbf{k}} - \varepsilon_{\mathbf{k}} \right],
  \\[1ex]
  \tilde{E} &=&
  -\, N \langle \hat{n}_{i}^{f} \rangle
  \left[ \tilde{\varepsilon}_{f} - \varepsilon_{f} \right] +
  \frac{D-1}{D} \langle \hat{n}_{i}^{f} \rangle \sum_{\mathbf{k}}
  \left[ \tilde{\varepsilon}_{\mathbf{k}} - \varepsilon_{\mathbf{k}} \right].
  \nonumber \\[-2ex]
  \label{G19}
  &&
\end{eqnarray}
As already mentioned above, the approximations (i), (ii) decouple the different
$\mathbf{k}$ values from each other so that Eq.~\eqref{G12} and \eqref{G13}
are completely similar to those obtained for the Fano-Anderson model (compare
Ref.~\cite{Hubsch}). Thus, two quasi-particle branches are obtained,
\begin{eqnarray}
  \label{G20}
  \tilde{\varepsilon}_{\mathbf{k}} &=&
  \frac{\tilde{\varepsilon}_{f} + \varepsilon_{\mathbf{k}}}{2} - 
  \frac{\mathrm{sgn}(\tilde{\varepsilon}_{f} - \varepsilon_{\mathbf{k}})}{2} 
  W_{\mathbf{k}}, \\
  \label{G21}
  \tilde{\omega}_{\mathbf{k}} & := &
  \tilde{\varepsilon}_{f} + D \tilde{\Delta}_{\mathbf{k}} \,=\,
  \frac{\tilde{\varepsilon}_{f} + \varepsilon_{\mathbf{k}}}{2} + 
  \frac{\mathrm{sgn}(\tilde{\varepsilon}_{f} - \varepsilon_{\mathbf{k}})}{2} 
  W_{\mathbf{k}},
\end{eqnarray}
where
\begin{eqnarray*}
  W_{\mathbf{k}} &=& 
  \sqrt{
    (\varepsilon_{\mathbf{k}} - \tilde{\varepsilon}_{f})^{2} + 
    4D|V_{\mathbf{k}}|^{2}
  }.
\end{eqnarray*}
Note that the one-particle energies \eqref{G20} and \eqref{G21} still depend
on two unknown quantities: the renormalized $f$ level
$\tilde{\varepsilon}_{f}$ and the $f$ occupation number 
$\langle \hat{n}_{i}^{f} \rangle$ (that determines $D$ as defined above).

In Ref.~\cite{Hubsch} all expectation values as well as the renormalized $f$
level $\tilde{\varepsilon}_{f}$ have been determined by functional
derivative of the free energy. However, here, this approach can not easily be
applied because the Hubbard operators contained in the renormalized
Hamiltonian \eqref{G17} do not fulfill the usual fermionic anti-commutator
relations. Furthermore, the derivation of the free energy would also lead to
problematic $\delta$ functions that are caused by the abrupt change of the
statistic of the quasi-particle excitations at $\tilde{\varepsilon}_{f}$. (In
Ref.~\cite{Hubsch} these contributions do not appear because both $c$-like
and $f$-like excitations are caused by fermionic quasi-particles.) Therefore,
a different approach has to be developed to determine the renormalized $f$
level $\tilde{\varepsilon}_{f}$ and the expectation values.

In the following, the expectation values of the original Hamiltonian
$\mathcal{H}$ will be calculated using the renormalized one-particle operators
as derived in 
Ref.~\cite{Hubsch},
\begin{eqnarray}
  \label{G22}
  c_{\mathbf{k}m}^{\dagger}(\lambda\rightarrow 0) &=&
  \tilde{u}_{\mathbf{k}} c_{\mathbf{k}m}^{\dagger} +
  \tilde{v}_{\mathbf{k}} \hat{f}_{\mathbf{k}m}^{\dagger}, \\
  \label{G23}
  \hat{f}_{\mathbf{k}m}^{\dagger}(\lambda\rightarrow 0) &=&
  - D \, \tilde{v}_{\mathbf{k}} c_{\mathbf{k}m}^{\dagger} +
  \tilde{u}_{\mathbf{k}} \hat{f}_{\mathbf{k}m}^{\dagger},
\end{eqnarray}
where we defined
\begin{eqnarray}
  \label{G24}
  \left| \tilde{u}_{\mathbf{k}} \right|^{2} &=&
  \frac{1}{2}
  \left\{
    1 - \frac{\varepsilon_{\bf k} - \tilde{\varepsilon}_{f}}{W_{\bf k}}
    \mathrm{sgn}\left( \tilde{\varepsilon}_{f} - \varepsilon_{\bf k} \right)
  \right\},\\
  \label{G25}
  \left| \tilde{v}_{\mathbf{k}} \right|^{2} &=&
  \frac{1}{2D}
  \left\{
    1 + \frac{\varepsilon_{\bf k} - \tilde{\varepsilon}_{f}}{W_{\bf k}}
    \mathrm{sgn}\left( \tilde{\varepsilon}_{f} - \varepsilon_{\bf k} \right)
  \right\}.
\end{eqnarray}
Thus, the required expectation values of the \textit{full} Hamiltonian
$\mathcal{H}$ can be traced back to those calculated with respect to the
\textit{renormalized} Hamiltonian $\tilde{\mathcal{H}}$ because 
$
  \langle A\rangle = 
  \lim_{\lambda\rightarrow 0}\langle A(\lambda) \rangle_{\mathcal{H}_{\lambda}}
$
holds,
\begin{eqnarray}
  \label{G26}
  \left\langle c^{\dagger}_{{\bf k}m} c_{{\bf k}m} \right\rangle
  &=&
  \frac{1}{2}\left[
    1 - \frac{\varepsilon_{\bf k} - \tilde{\varepsilon}_{f}}{W_{\mathbf{k}}}
    \mathrm{sgn}\left( \tilde{\varepsilon}_{f} - \varepsilon_{\bf k} \right)
  \right] 
  f(\tilde{\varepsilon}_{\mathbf{k}}) \\
  &&
  + \, \frac{1}{2}\left[
    1 + \frac{\varepsilon_{\bf k} - \tilde{\varepsilon}_{f}}{W_{\mathbf{k}}}
    \mathrm{sgn}\left( \tilde{\varepsilon}_{f} - \varepsilon_{\bf k} \right)
  \right] 
  \bar{f}(\tilde{\omega}_{\mathbf{k}}), \nonumber
\end{eqnarray}
\begin{eqnarray}
  \label{G27}
  \lefteqn{
    \left\langle
      \hat{f}^{\dagger}_{{\bf k}m} c_{{\bf k}m} + \mathrm{h.c.}
    \right\rangle
    \, = \,
  } && \\
  &=&
  - \, 2 \,
  \mathrm{sgn}\left( \tilde{\varepsilon}_{f} - \varepsilon_{\bf k} \right)
  \frac{D|V_{\mathbf{k}}|}{W_{\mathbf{k}}}
  \left[
    f(\tilde{\varepsilon}_{\mathbf{k}}) - 
    \bar{f}(\tilde{\omega}_{\mathbf{k}})
  \right].
  \nonumber
\end{eqnarray}
Here, we introduced the Fermi function
\begin{eqnarray*}
  f(\tilde{\varepsilon}_{\mathbf{k}}) & := & 
  \langle  c^{\dagger}_{{\bf k}m} c_{{\bf k}m} \rangle_{\tilde{\mathcal{H}}} =
  \frac{1}{1 + e^{\beta\tilde{\varepsilon}_{\mathbf{k}}}},
\end{eqnarray*}
and defined
\begin{eqnarray}
  \label{G28}
  \bar{f}(\tilde{\omega}_{\mathbf{k}}) & := & 
  \frac{1}{D} \langle
    \hat{f}^{\dagger}_{{\bf k}m} \hat{f}_{{\bf k}m} 
  \rangle_{\tilde{\mathcal{H}}}  .
\end{eqnarray}
Note that the factor $D$ in Eq.~\eqref{G28} has been introduced to underline
the similarities of Eq.~\eqref{G26} and \eqref{G27} with the corresponding
results of the analytical treatment of Ref.~\cite{Hubsch}.

In principle, the $f$ occupation number $\langle \hat{n}_{i}^{f} \rangle$
could also be calculated using the renormalized one-particle
operators. However, here we alternatively employ the particle conservation
under unitary transformations. Thus, we obtain
\begin{eqnarray}
  \label{G29}
  \lefteqn{\langle \hat{n}_{i}^{f} \rangle \,=\, } && \\
  &=&
  \frac{1}{2} \frac{\nu_{f}}{N} \sum_{\mathbf{k}}
  \left[
    1 + \frac{\varepsilon_{\bf k} - \tilde{\varepsilon}_{f}}{W_{\mathbf{k}}}
    \mathrm{sgn}\left( \tilde{\varepsilon}_{f} - \varepsilon_{\bf k} \right)
  \right]
  f(\tilde{\varepsilon}_{\mathbf{k}}) \nonumber \\
  && + \, 
  \frac{1}{2} \frac{\nu_{f}}{N} \sum_{\mathbf{k}}
  \left[
    2D - 1 - 
    \frac{\varepsilon_{\bf k} - \tilde{\varepsilon}_{f}}{W_{\mathbf{k}}}
    \mathrm{sgn}\left( \tilde{\varepsilon}_{f} - \varepsilon_{\bf k} \right)
  \right]
  \bar{f}(\tilde{\omega}_{\mathbf{k}}). \nonumber
\end{eqnarray}

For actual calculations one needs to evaluate Eq.~\eqref{G28} in order to
determine the expectation values of the full Hamiltonian as given in
Eqs.~\eqref{G26}, \eqref{G27}, and \eqref{G29}. Because of the unusual
properties of the Hubbard operators, there is no straightforward way to
evaluate Eq.~\eqref{G28} and further approximations are necessary. As long as
the renormalized $f$ level is situated \textit{above} the chemical potential a
mean-field treatment of the electronic correlations contained in $\mathcal{H}$
might be sufficient, and we would find
$\bar{f}(\tilde{\omega}_{\mathbf{k}})\approx f(\tilde{\omega}_{\mathbf{k}})$
as directly obtained by employing approximation (iii) mentioned above. On the
other hand, here we are also interested in solutions of the PAM with a
renormalized $f$ level \textit{below} the Fermi level which require a
theoretical treatment of the electronic correlations in $\mathcal{H}$
\textit{beyond} a mean-field approximation. Therefore, Eq.~\eqref{G28} is
evaluated as follows
\begin{eqnarray*}
  \langle 
    \hat{f}^{\dagger}_{{\bf k}m} \hat{f}_{{\bf k}m} 
  \rangle_{\tilde{\mathcal{H}}}
  &=&
  \frac{1}{\mathrm{Tr}\, e^{-\beta \tilde{\mathcal{H}}}}
  \mathrm{Tr}\left(
    e^{\beta \tilde{\mathcal{H}}} \hat{f}_{{\bf k}m}
    e^{-\beta \tilde{\mathcal{H}}}
    \hat{f}^{\dagger}_{{\bf k}m}
    e^{-\beta \tilde{\mathcal{H}}}
  \right) \\
  &\approx&
  f(\tilde{\omega}_{\mathbf{k}})
  \left\langle 
    \left\{
      \hat{f}^{\dagger}_{{\bf k}m}, \hat{f}_{{\bf k}m} 
    \right\}_{+}
  \right\rangle_{\tilde{\mathcal{H}}}
\end{eqnarray*}
where the approximated $f$ excitation energy as derived in Eq.~\eqref{G10} has
been used. Thus, \eqref{G28} can be rewritten as 
\begin{eqnarray}
  \label{G30}
  \bar{f}(\tilde{\omega}_{\mathbf{k}}) & := & 
  \frac{\frac{1}{D} f(\tilde{\omega}_{\mathbf{k}})}
  {
    1 + \frac{\nu_{f} - 1}{N}\sum_{\mathbf{k}'} f(\tilde{\omega}_{\mathbf{k}'})
  } .
\end{eqnarray} 
Unfortunately, approximation \eqref{G30} does not offer a direct link to the
mean-field result, 
$\bar{f}(\tilde{\omega}_{\mathbf{k}})\approx f(\tilde{\omega}_{\mathbf{k}})$,
for renormalized $f$ energies above the Fermi level. Thus, differences between
the presented treatment and the analytical solution of Ref.~\cite{Hubsch} will
appear. 

\bigskip
At this point all physical quantities can be calculated as function of the
renormalized $f$ energy $\tilde{\varepsilon}_{f}$. Because we have employed
approximation (ii) it is not possible anymore to use the renormalization
equation \eqref{G15} for $\mu_{f,\lambda}$ to determine
$\tilde{\varepsilon}_{f}$. Therefore, the local $f$ energy
$\tilde{\varepsilon}_{f}$ is considered as a free parameter and is determined
by minimization of the free energy. Because of the unusual anticommutator
relations of the Hubbard operators $\hat{f}^{\dagger} _{\mathbf{k}m}$, the
free energy can not be directly determined. Instead, 
\begin{eqnarray}
  \frac{\mbox{d}F}{\mbox{d}\tilde{\varepsilon}_{f}} &=&
   \sum _{{\bf k},m} 
   \frac{
     \mbox{d}\tilde{\varepsilon}_{{\bf k}}
   }{\mbox{d}\tilde{\varepsilon}_{f}}
  \left\langle
    c^{\dagger}_{{\bf k}m} c_{{\bf k}m}
  \right\rangle_{\tilde{\mathcal{H}}}
  + 
  \frac{\mbox{d}\tilde{\mu}_{f}}{\mbox{d}\tilde{\varepsilon}_{f}}
  \sum _{\mathbf{k},m} 
  \left\langle 
    \hat{f}^{\dagger}_{\mathbf{k}m} \hat{f}_{\mathbf{k}m} 
  \right\rangle_{\tilde{\mathcal{H}}} \nonumber\\
  \label{G31}
  &&
  + \sum _{{\bf k},m} 
  \frac{\mbox{d}\tilde{\Delta}_{\mathbf{k}}}{\mbox{d}\tilde{\varepsilon}_{f}}
  \left\langle \left(
    \hat{f}^{\dagger} _{\mathbf{k}m} \hat{f}_{\mathbf{k}m} 
  \right)_{\mathrm{NL}} \right\rangle_{\tilde{\mathcal{H}}}
  + \frac{\mbox{d}\tilde{E}}{\mbox{d}\tilde{\varepsilon}_{f}}
\end{eqnarray}
is numerically integrated in order to calculate the free energy $F$ as
function of the renormalized $f$ energy $\tilde{\varepsilon}_{f}$. Note that
Eq.~\eqref{G31} has been obtained from the renormalized Hamiltonian
\eqref{G17}. Actual results are discussed in the next section.

\begin{figure}
\begin{center}
  \scalebox{0.6}{
    \includegraphics*{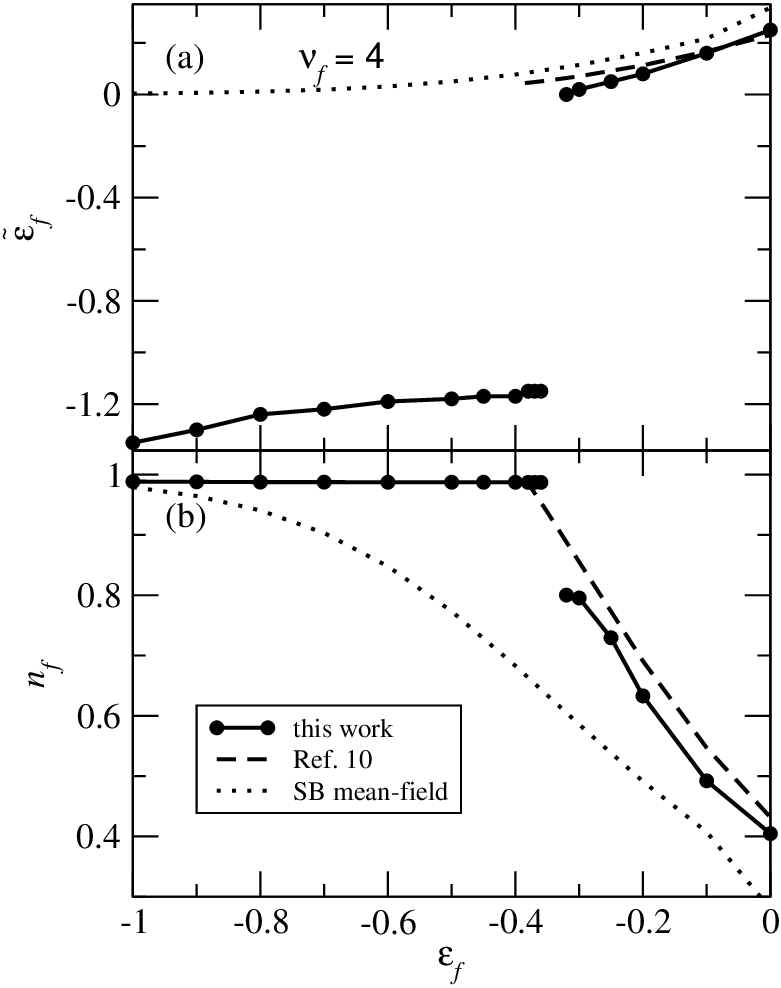}
  }
\end{center}
\caption{
  Renormalized $f$ level $\tilde{\varepsilon}_{f}$ [panel (a)] and averaged
  $f$ occupation number  $n_{f} = \langle \hat{n}_{i}^{f} \rangle$ [panel (b)]
  as function of the unrenormalized $f$ energy $\varepsilon_{f}$ where an
  one-dimensional PAM ($N=10000$, $\nu_{f}=4$, $\nu_{f}V^{2} = 0.36$, $\mu=0$,
  $T=0.00001$) with a linear dispersion relation for the conduction band in
  the energy range between $-1$ and $1$ has been considered. (All energies are
  given in units of the half bandwidth.) For comparison, the results of the
  PRM approach of Ref.~\cite{Hubsch} and of the SB mean-field theory are drawn
  with dashed and dotted lines.
}
\label{Fig_1}
\end{figure}

\begin{figure}
\begin{center}
  \scalebox{0.6}{
    \includegraphics*{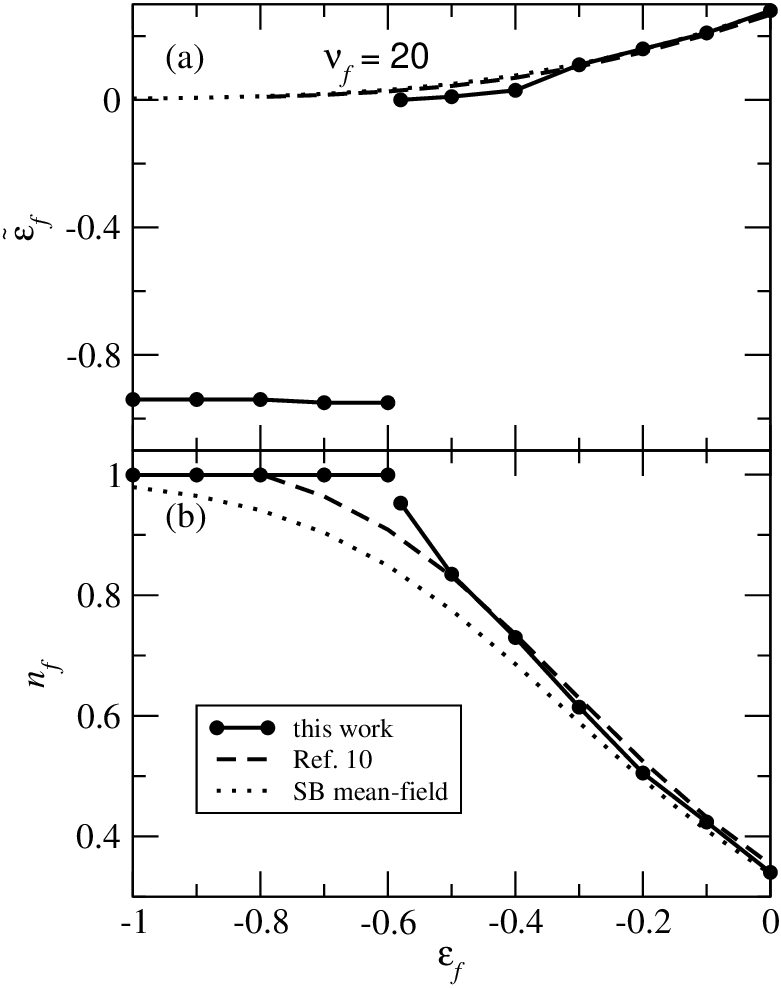}
  }
\end{center}
\caption{
  Renormalized $f$ level $\tilde{\varepsilon}_{f}$ [panel (a)] and averaged
  $f$ occupation number  $n_{f} = \langle \hat{n}_{i}^{f} \rangle$ [panel (b)]
  as function of the unrenormalized $f$ energy $\varepsilon_{f}$ for an
  one-dimensional PAM with $\nu_{f}=20$. Other parameters are chosen as in
  Fig.~\ref{Fig_1}.
}
\label{Fig_2}
\end{figure}

\section{Results}
\label{Results}

It is believed that the one-particle energy $\varepsilon_{f}$ of the localized
\textit{f} electrons is smoothly changed in CeCu$_{2}$Si$_{2}$ due to
pressure \cite{Holmes}. Therefore, we want to discuss the physical properties
of the PAM as a function of $\varepsilon_{f}$. 

\bigskip
At first let us consider an one-dimensional PAM with a linear dispersion
relation $\varepsilon_{\mathbf{k}}$ for the conduction electrons in the energy
range between $-1$ and $1$, and a $\mathbf{k}$ independent hybridization
$V_{\mathbf{k}} = V$. The other parameters are chosen as follows
$\nu_{f}V^{2}=0.36$, chemical potential $\mu=0$, and $T=0.00001$ where all 
energies are given in units of the half bandwidth. 

As one can see from Figs.~\ref{Fig_1} and \ref{Fig_2}, we obtain two different 
types of solutions depending on the value of the unrenormalized $f$ level
$\varepsilon_{f}$. First of all, we obtain the usual SB type solutions with
intermediate valence states $n_{f}<1$ where the renormalized energy
$\tilde{\varepsilon}_{f}$ is energetically located above the Fermi energy. If
the  unrenormalized energy $\varepsilon_{f}$ is lowered the renormalization
contributions are no longer sufficient to push $\tilde{\varepsilon}_{f}$
above the Fermi level, and the renormalized $f$ energy
$\tilde{\varepsilon}_{f}$ is located far below the Fermi energy. In this case, 
the averaged \textit{f} occupation $n_{f}$ is almost exactly 1 and an integral
valence state is obtained.

Figs.~\ref{Fig_1} and \ref{Fig_2} also reveal the very good agreement between
the HF type solutions of the presented PRM approach and the analytical results
of Ref.~\cite{Hubsch}. In this way it is proven that the Hubbard operators
can be replaced by usual fermionic operators (compare approximation (iii) in
Sec.~\ref{Approx_Sol}) because in this case only very few $f$ type states
below the Fermi level are occupied as discussed above. In this regard one
needs to keep in mind an important difference between the SB theory and our PRM
approach: The quasi-particles of the SB theory change their character as
function of $\mathbf{k}$ between more \textit{f}-like and more \textit{c}-like
behavior. In the PRM excitations do not change their character 
as function of $\mathbf{k}$, and the quasi-particle energies show jumps in
their $\mathbf{k}$ dependence if $\tilde{\varepsilon}_{f}$ is energetically
located within the conduction band. Note, however, that the various parts of
the quasiparticle bands fit perfectly together, as one can see from
Eqs.~\eqref{G20} and \eqref{G21}. 

For comparison, the results of the analytical solution of Ref.~\cite{Hubsch}
and of the SB theory are shown as well in Figs.~\ref{Fig_1} and
\ref{Fig_2}. As one can see, no solution with renormalized \textit{f} level
$\tilde{\varepsilon}_{f}<0$ could be found for these analytical approaches
because both do not explicitly ensure that \textit{f} sites can only be either
empty or singly occupied as already discussed above. 

\bigskip
The well-defined transition between the two different solution types is of
particular interest. As expected, for the HF-like solution the
\textit{f}-charge is always smaller than 1 due to hybridization processes
between \textit{f} and \textit{c}-electrons. Simultaneously heavy
quasiparticle bands are formed at the Fermi surface. To describe the HF
behavior the full Anderson model has to be considered. As the bare
\textit{f}-level moves to smaller energies a transition to an integral valence
charge of $n_{f}=1$ is observed (similar to the Anderson impurity model
\cite{Haldane}). In this case only the \textit{c} electrons should form the
Fermi surface. Thus, the observed valence transition can also be interpreted
as a collapse of the large Fermi surface of the HF state which is formed by
conduction as well as by localized \textit{f} electrons. Note, however, that
the question whether localized electrons contribute to the Fermi sea volume or
not is still controversially discussed in the literature \cite{Oshikawa}.

As one can see from Figs.~\ref{Fig_1} and \ref{Fig_2}, the obtained valence
transition is much more pronounced for small degeneracies $\nu_{f}$, and a
smooth transition can be expected in the limit $\nu_{f}\rightarrow\infty$ of
the SB theory. Therefore, a sharp valence change in generalized SB theories
can only be obtained if a rather large additional Coulomb repulsion between
\textit{f} and conduction electrons is present in the system 
\cite{Holmes,Onishi}. However, here we have shown that such a valence
transition can also be obtained in the plain PAM if corrections for small
degeneracies $\nu_{f}$ are properly taken into account. 

\begin{figure}
\begin{center}
  \scalebox{0.59}{
    \includegraphics*{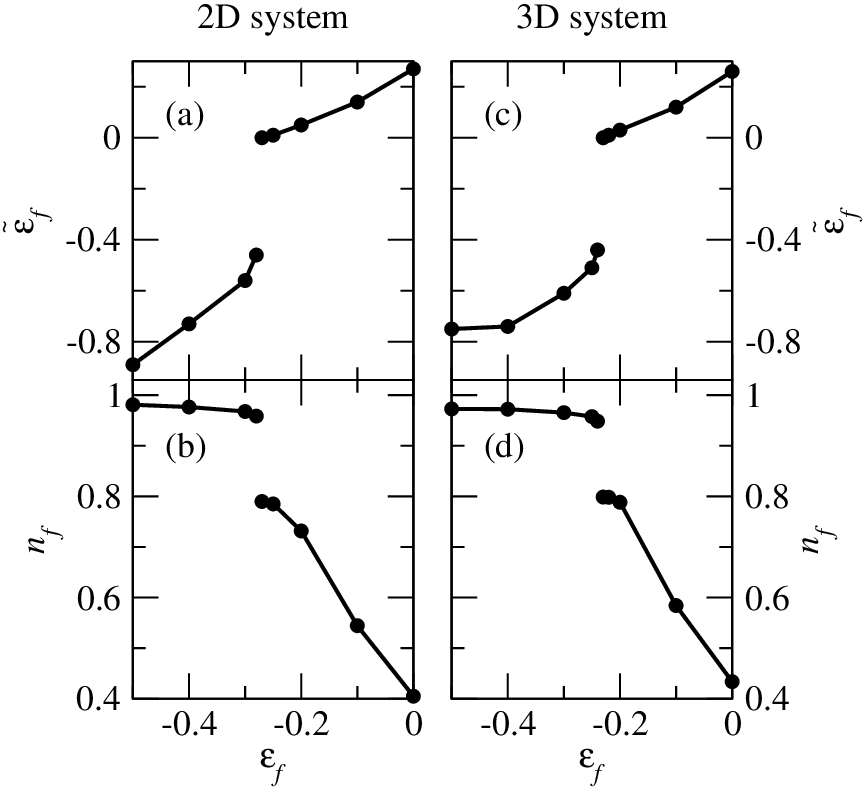}
  }
\end{center}
\caption{
  Panel (a) and (b) [(c) and (d)] show the results for the renormalized
  \textit{f} level $\tilde{\varepsilon}_{f}$ and the \textit{f} occupation
  number $n_{f}=\langle \hat{n}_{i}^{f} \rangle$ for a two-dimensional
  [three-dimensional] system with $100\times 100$ [$26\times26\times26$] 
  lattice   sites. As in Figs.~\ref{Fig_1} and \ref{Fig_2}, a linear
  dispersion relation $\varepsilon_{\mathbf{k}} = \varepsilon(|\mathbf{k}|)$
  has been chosen, and the electronic band covers an energy range between -1.5
  and 0.5 [-1.8 and 0.2] where $\mu=0$, $\nu_{f}=4$, $\nu_{f}V^{2}=0.36$, and
  $T=0.00001$. (Energies are given in units of the half bandwidth.) Note that
  the conduction band has been energetically shifted in in order to ensure a
  filling comparable to the one-dimensional case of Fig.~\ref{Fig_1} because a
  smaller filling of the conduction band reduces the change in the $f$
  occupation $n_{f}$ at the transition point.
}
\label{Fig_3}
\end{figure}

\bigskip
One of the advantages of the analytical PRM is the opportunity to consider
much larger systems than accessible by numerical methods. Therefore, we are 
also able to study two- and three-dimensional systems of reasonable sizes. In
this way we can easily show that the observed valence transition is not an
unique phenomenon of the one-dimensional PAM. The valence transition also
occurs in two- and three-dimensional systems as can been seen in
Fig.~\ref{Fig_3}. Therefore, the observed behavior has to be considered as a
general feature of the PAM, and our results should also be of relevance for
actual physical HF systems like CeCu$_2$Si$_2$ or related compounds.

\section{Discussion and Summary}
\label{Summ}

The occurrence of a valence transition in the plain PAM is the main finding of
this paper. In contrast, a rather large additional Coulomb repulsion has been
claimed to be necessary for the valence transition in an extended PAM 
\cite{Holmes,Onishi}. The studies of Refs.~\cite{Holmes,Onishi} were based on
a slave-boson fluctuation approximation that extends the well-known
slave-boson mean-field theory \cite{Coleman,Fulde} but still employs the limit
of large degeneracy $\nu_{f}\rightarrow\infty$. Our results show (compare
Figs.~\ref{Fig_1} and \ref{Fig_2}) that the observed valence transition
becomes smooth in this limit. Therefore, it is reasonable that an additional
interaction was found to be necessary in order to obtain a valence transition
in an approach employing  $\nu_{f}\rightarrow\infty$.

Our work also shows the importance of taking care of a physical $f$ occupation
in theoretical approaches. In particular, it turns out that a completely
uncorrelated model is not able to prevent from unphysical multiple
occupation of \textit{f} sites, and no integral valence states can be found in
this way. In contrast, the presented PRM approach to the PAM explicitly
suppresses unphysical multiple \textit{f} occupation which is, in particular,
crucial for integral valence states. 

We obtain two solution types: a mixed valence state with a renormalized
\textit{f} level $\tilde{\varepsilon}_{f}$ above the Fermi energy and an
integral valence state with $\tilde{\varepsilon}_{f}$ below the Fermi
level. Furthermore, parameter regimes exist where the transition between the
two solution types is accompanied by a drastic change in the \textit{f}
occupation. Such a sharp valence transition occurs in one-dimensional as well
as in two- and three-dimensional systems so that this behavior has to be
considered as a general feature of the PAM. Note that a similar valence
transition has been experimentally found in CeCu$_2$Si$_2$ from high pressure
experiments \cite{Yuan,Holmes}. 

In the case of an integral valence state one would expect that the system can
be described by a Kondo Hamiltonian which is gained from the PAM by the
Schrieffer-Wolff transformation \cite{Schrieffer} for 
$V/|\varepsilon_f|\ll 1$. Note, however, that in the present approach 
spin fluctuations have been neglected altogether but Kondo-like and 
RKKY-like interactions as well as higher charge fluctuation terms 
are automatically generated during the renormalization procedure. 
These contributions will have to be considered in the future. 
One might expect that additional spin and charge fluctuations 
might possibly give rise to magnetic and superconducting phases both for the
intermediate valence and for the integer valence regime. Also, one may
speculate that the magnitude of the magnetic moment will be different for
these cases due to additional screening processes.

\bigskip
The PRM approach presented in this paper only addresses the question for the
valence transition in the plain PAM. However, as mentioned above, the PRM
scheme offers great opportunities to include additional interactions which are 
automatically generated during the renormalization procedure. Therefore,
extensions of the PRM treatment might be promising starting points to study
the competing interactions in CeCu$_2$Si$_2$ and related compounds in more
detail.

\bigskip
\textit{Acknowledgment.}
We would like to acknowledge stimulating and enlightening discussions with
T. Bryant and V. Zlati{\'c}.  This work was supported by the DFG through the
research program SFB 463. AH is grateful for the support of the DFG through
Grant No. HU~993/1-1, and of the US Department of Energy, Division of
Materials Research, Office of Basic Energy Science.

{}

\end{document}